\documentclass[3p]{elsarticle}
\usepackage{amsmath}    % need for subequations
\usepackage{graphicx}   % need for figures
\usepackage{color}
\usepackage{hyperref}   % use for hypertext links, including those to external documents and URLs

\newcommand{\nuc}[2]{\textsuperscript{#1}#2}

\journal{Nuclear Physics A}

\begin{document}
    \begin{frontmatter}
        \title{Observation of a low-lying neutron-unbound state in \nuc{19}{C}}
    %%%%%%%%%%%%%%%%%
    %% AUTHOR LIST %%
    %%%%%%%%%%%%%%%%%
    \author[nscl,msu]{M.~Thoennessen\corref{cor1}}
    \ead{thoennessen@nscl.msu.edu}
    \author[nscl,msu]{S.~Mosby\fnref{smaddy}}
    \author[rhodes]{N.~S.~Badger}
   \author[nscl]{T.~Baumann}
   \author[nscl]{D.~Bazin}
    \author[westmont]{M.~Bennett}
    \author[wabash]{J.~Brown}
    \author[nscl,msu]{G.~Christian\fnref{gcaddy}}
    \author[hope]{P.~A.~DeYoung}
    \author[cmu]{J.~E.~Finck}
    \author[westmont]{M.~Gardner}
%    \author[iusb]{J.~D.~Hinnefeld}
    \author[rhodes]{E.~A.~Hook}
    \author[concordia]{B.~Luther}
    \author[rhodes]{D.~A.~Meyer}
    \author[nscl,msu]{M.~Mosby}
%    \author[hope]{G.~F.~Peaslee}
    \author[westmont]{W.~F.~Rogers}
    \author[nscl,msu]{J.~K.~Smith}
    \author[nscl,msu]{A.~Spyrou}
    \author[nscl,msu]{M.~J.~Strongman}

    \cortext[cor1]{Corresponding Author.}
    \fntext[smaddy]{Present Address: LANL, Los Alamos, NM 87545, USA}
    \fntext[gcaddy]{Present Address: TRIUMF, 4004 Wesbrook Mall, Vancouver, British Columbia V6T 2A3, Canada}
    %%%%%%%%%%%%%%%%%%
    %% ADDRESS LIST %%
    %%%%%%%%%%%%%%%%%%
    \address[nscl]{National Superconducting Cyclotron Laboratory, Michigan State University, East Lansing, Michigan 48824}
    \address[msu]{Department of Physics \& Astronomy, Michigan State University, East Lansing, Michigan 48824}
    \address[rhodes]{Department of Physics, Rhodes College, Memphis, Tennessee 38112}
    \address[westmont]{Department of Physics, Westmont College, Santa Barbara, California 93108}
    \address[wabash]{Department of Physics, Wabash College, Crawfordsville, Indiana 47933}
    \address[hope]{Department of Physics, Hope College, Holland, Michigan 49423}
    \address[cmu]{Department of Physics, Central Michigan University, Mt. Pleasant, Michigan, 48859}
    \address[iusb]{Department of Physics \& Astronomy, Indiana University at South Bend, South Bend, Indiana 46634}
    \address[concordia]{Department of Physics, Concordia College, Moorhead, Minnesota 56562}

    \begin{abstract}

     Proton removal reactions from a secondary \nuc{22}{N} beam were utilized to populate unbound states in neutron-rich carbon isotopes. Neutrons were measured with the Modular Neutron Array (MoNA) in coincidence with carbon fragments. A resonance with a decay energy of 76(14)~keV was observed in the system \nuc{18}{C} $+ n$ corresponding to a state in \nuc{19}{C} at an excitation energy of 653(95)~keV. This resonance could correspond to the first $5/2^+$ state which was recently speculated to be unbound in order to describe 1n and 2n removal cross section measurements from \nuc{20}{C}.

    \end{abstract}


    \end{frontmatter}

    \section{Introduction}

Neutron decay spectroscopy has been used extensively to study excited unbound states in neutron-rich oxygen isotopes \cite{Elekes2007,Schiller2007,Hoffman2009,Hoffman2011,Satou2012,Tsoo2012,Lapoux2012} close to the dripline and even unbound nuclei beyond the dripline \cite{Hoffman2008,Lunderberg2012}. These studies were instrumental in establishing the emergence of the $N$ = 14 and $N$ = 16 subshell closures along the $Z$ = 8 line.  The evolution of these gaps towards lighter-$Z$ nuclei is less clear and decay properties of neutron-unbound states have only been measured in \nuc{17}{C} and \nuc{19}{C} \cite{Satou2008}. 

\nuc{19}{C} was first observed in 1974 by Bowman \textit{et al.} \cite{Bowman1974,Thoennessen2012} and it is the first bound nuclide in the $A =3Z+1$ sequence. All lighter nuclides within this series, \nuc{4}{H},  \nuc{7}{He},  \nuc{10}{Li},  \nuc{13}{Be}, and \nuc{16}{B} are unbound. While the spin and parity of the ground state has been established to be $1/2^+$ \cite{Nakamura1999,Maddalena2001}, the level structure of the excited states is still debated \cite{Yuan2012}. A first bound excited state was reported in 2004 in a fragmentation reaction at an excitation energy of 201(15)~keV \cite{Stanoiu2004,Stanoiu2008} and was later confirmed in a proton inelastic scattering experiment at 197(6)~keV \cite{Elekes2005}. The latter experiment observed an additional $\gamma$-ray coincident with an energy of 72(4)~keV which placed a second bound excited state at an excitation energy of 269(8)~keV. These states were tentatively assigned spin and parities of $3/2^+$ and $5/2^+$, respectively  \cite{Elekes2005}. However, recently one- and two-neutron removal experiments questioned the assignment of the second excited state, where the authors of the papers argued that the measured cross sections could only be explained if the $5/2^+$ state were unbound \cite{Ozawa2011,Kobayashi2012}. 
Presently, only one unbound state at a fairly high excitation energy of 1.46(10)~MeV is known and has been assigned to be the second excited $5/2^+$ state \cite{Satou2008}. The calculated spectroscopic factor of this state is small and thus cannot account for the measured neutron removal cross sections.

We performed a neutron decay spectroscopy experiment measuring unbound states in neutron-rich carbon isotopes via the one-proton (plus two neutron) removal reaction from a \nuc{22}{N} beam to search for low-lying unbound states in \nuc{19}{C}.

    \section{Experimental Setup and Data Analysis}

The experiment was performed at the National Superconducting Cyclotron Laboratory (NSCL) at Michigan State University. A 90~pnA \nuc{48}{Ca} primary beam at 140 MeV/u impinged on a 2068 mg/cm\textsuperscript{2} \nuc{9}{Be} production target, and isotopic separation of a 68 MeV/u \nuc{22}{N} secondary beam was achieved using the A1900 fragment separator \cite{Morrissey2003} with a 1057 mg/cm\textsuperscript{2} Al achromatic wedge degrader placed at the dispersive image. All data were taken with A1900 momentum slits set at 2.5\% acceptance, which resulted in a \nuc{22}{N} particle rate of 37/s with a purity of 32\%. The main contaminants were light ions with \nuc{26}{F} (6\%) and \nuc{20}{C} (2.8\%) being the largest contributions of heavy ions \cite{Mosby2012}. The secondary beam bombarded a 481~mg/cm$^2$ $^9$Be reaction target in front of a large-gap superconducting dipole magnet \cite{Bird2005} which bent the beam and charged fragments from the reactions away from the beam axis. Neutrons from the reactions were detected in coincidence near zero degrees by the Modular Neutron Array (MoNA) \cite{Baumann2005}.

\begin{figure}
        \begin{center}
                \includegraphics[width=0.6\textwidth]{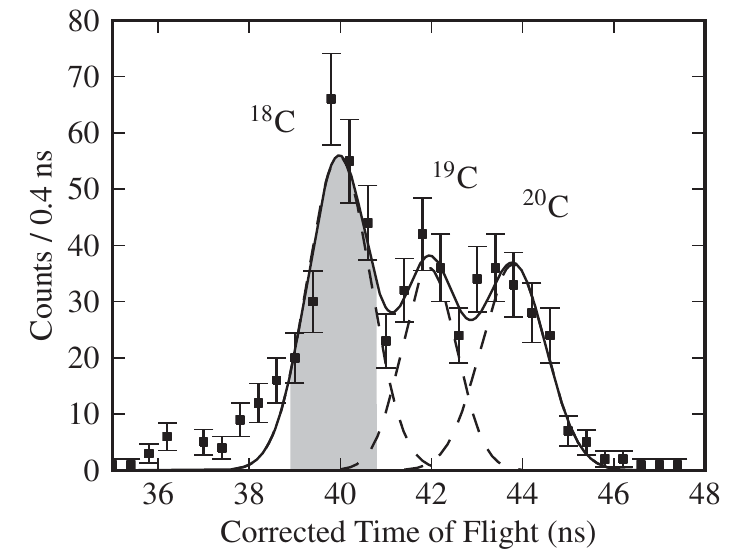}
        \end{center}
        \caption{Identification of the detected carbon isotopes. The selection of the \nuc{18}{C} events used for the decay energy spectrum of \nuc{19}{C} is indicated by the grey shading.}
        \label{fig:eid}
    \end{figure}

Isotopic identification was achieved by using energy loss, time of flight, and angle/position correlation information following the procedure described in Ref. \cite{Christian2012}. The identified carbon fragments following one-proton (or 1p1n and 1p2n) removal reactions are shown in Figure~\ref{fig:eid} where the selected \nuc{18}{C} fragments from the decay of \nuc{19}{C} are indicated by the grey area. The contribution of \nuc{19}{C} in the \nuc{18}{C} gate was about 2\%. Coincidences of neutrons with \nuc{20}{C} were also analyzed and have already been published separately where further details of the experimental setup and data analysis are described \cite{Mosby2012}.

The decay energy of \nuc{19}{C} was reconstructed using the invariant mass method with the momentum vectors of the charged fragments and the neutrons measured in coincidence.  Figure \ref{fig:edecay} shows the decay spectrum which displays a sharp peak below 200~keV. Monte Carlo simulations were performed to extract the resonance parameters of this peak. The simulations included the full experimental setup, reaction process, and decay characteristics. Further details can be found in Ref.~\cite{Mosby2012}. The near-threshold resonance was modeled using a Breit-Wigner lineshape with an energy dependent width. Although \nuc{19}{C} can be produced directly in a 1p2n removal reaction it also can be populated following the emission of continuum neutrons from \nuc{21}{C} and \nuc{20}{C}. The contribution of these continuum neutrons was included in the simulation using a Maxwellian distribution.

\section{Results and Discussion}

\begin{figure}[t]
        \begin{center}
            \includegraphics[width=0.8\textwidth]{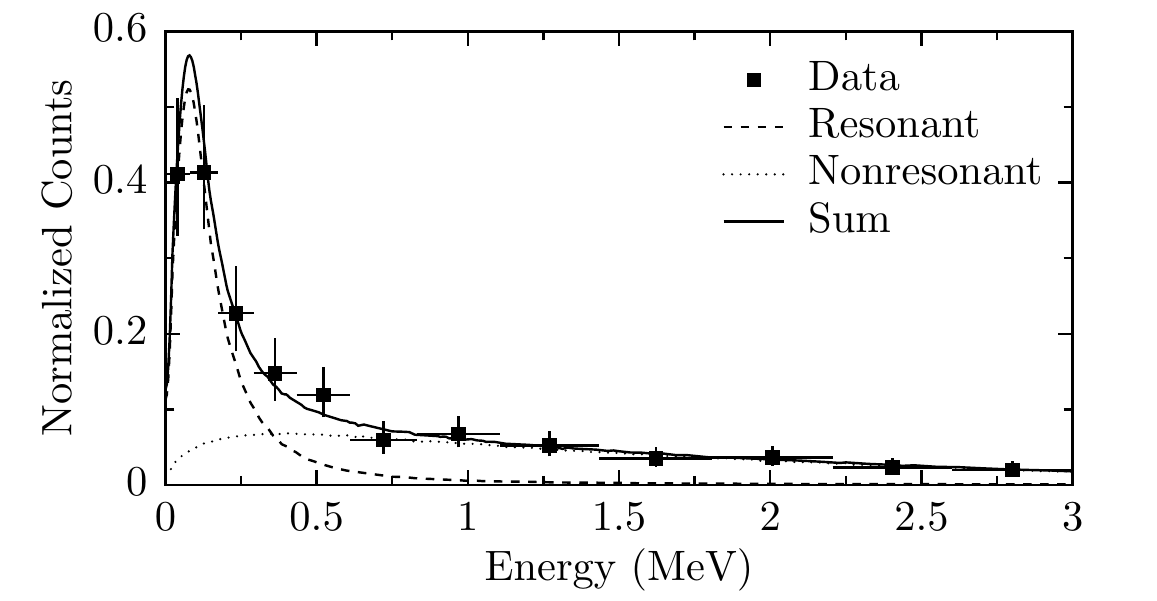}
        \end{center}
        \caption{Reconstructed decay energy spectra for \nuc{19}{C}. The data are compared with simulations (solid line) which include contributions from a resonance at 76(14)~keV (dashed line) and a non-resonant background (dotted line).}
        \label{fig:edecay}
    \end{figure}

The best fit to the data (solid line in Figure \ref{fig:edecay}) was found to result from a resonance with a decay energy of 76(14)~keV (dashed line) and a Maxwellian non-resonant background (dotted line) with a temperature parameter of 2.28~MeV. The resonance contribution to the total yield was 54\% corresponding to a cross section of 0.8(2)~mb. This is consistent with an upper limit of the direct 1p2n removal cross section estimated within the Eikonal method \cite{Tostevin1999,Bazin2003} of 0.32~mb. As mentioned earlier, the additional cross section to populate this resonance could originate from other evaporation-type mechanisms. The fit was not sensitive to the width parameter as it is dominated by the experimental resolution of 100~keV \cite{Mosby2012}. This  exceeds the calculated single-particle width of 10~keV which was used in the calculations shown in Figure \ref{fig:edecay}.

The excitation energy of the resonance is equal to the sum of the resonance energy and the one-neutron separation energy of \nuc{19}{C}. The most recent mass evaluation quotes a value of 580(90)~keV \cite{Wang2012} which is unchanged from the 2003 value \cite{Audi2003}, and corresponds to the rounded value of 576.83~keV with an uncertainty of 93.70~keV quoted in the online data base \cite{Mass2003}. Thus the excitation energy of the observed resonance is 653(95)~keV.

The data show no evidence for the presence of the previously reported state at 1.46(10)~MeV  \cite{Satou2008} which would correspond to a decay energy of 880~keV. At this decay energy, the  efficiency is reduced by a factor of two and the width of a resonance is increased by a factor of three relative to the observed peak at 76~keV (see Figure 6 of Ref.~\cite{Mosby2012}). Assuming about equal populations of the two states, simulations indicated that the 880~keV peak is approximately equal to the number of counts observed in this energy region and not distinguishable from a non-resonant background.

There is a plausible explanation as to why the proton inelastic scattering experiment did not observe the presently reported low-lying resonance. Assuming that both states have a spin and parity of $5/2^+$, the calculated (p,p') cross section (using shell model wave functions with the WBP interaction) for populating the second $5/2^+_2$ state is about a factor of 10 larger than that for populating the first $5/2^+_1$ state. Thus it would not have been observable above the measured background near zero decay energy \cite{Satou2008}. This selective observation of these two states can be understood from the two different reaction mechanisms -- proton removal reactions in the present experiment and proton inelastic scattering in Ref. \cite{Satou2008} -- used to populate the states.

The presence of an unbound state close to threshold has recently been postulated in two separate measurements of one- and two-neutron removal reactions. Ozawa \textit{et al.} bombarded a liquid hydrogen target with \nuc{20}{C} at 40 MeV/nucleon \cite{Ozawa2011} and Kobayashi \textit{et al.} used 240 MeV/nucleon \nuc{20}C on a carbon target \cite{Kobayashi2012}. In the first experiment the two-neutron removal cross section was larger while the one-neutron removal cross section was smaller than results of eikonal model calculations. Ozawa \textit{et al.} \cite{Ozawa2011} could resolve this discrepancy by suggesting that the first $5/2^+_1$ state in \nuc{19}{C} is unbound: ``if the order of the level with E$_x$ = 0.19~MeV and that with E$_x$ = 0.62 MeV swaps and the first $5/2^+$ state becomes unbound, theoretical cross sections will approach the experimental ones... Thus, if the first $5/2^+$ state is unbound, it may be located just above the neutron decay threshold and decay by a low-energy neutron that may have been too weak to observe in Ref. \cite{Satou2008}.'' Kobayashi \textit{et al.} \cite{Kobayashi2012} could reproduce both cross sections well by assuming that the $5/2^+_1$ was unbound: ``There is no evidence from this work that the $5/2^+_1$ shell-model state in \nuc{19}{C} is bound. It is assumed to be unbound.''

    \begin{figure}[t]
        \begin{center}
            \includegraphics[width=0.95\textwidth]{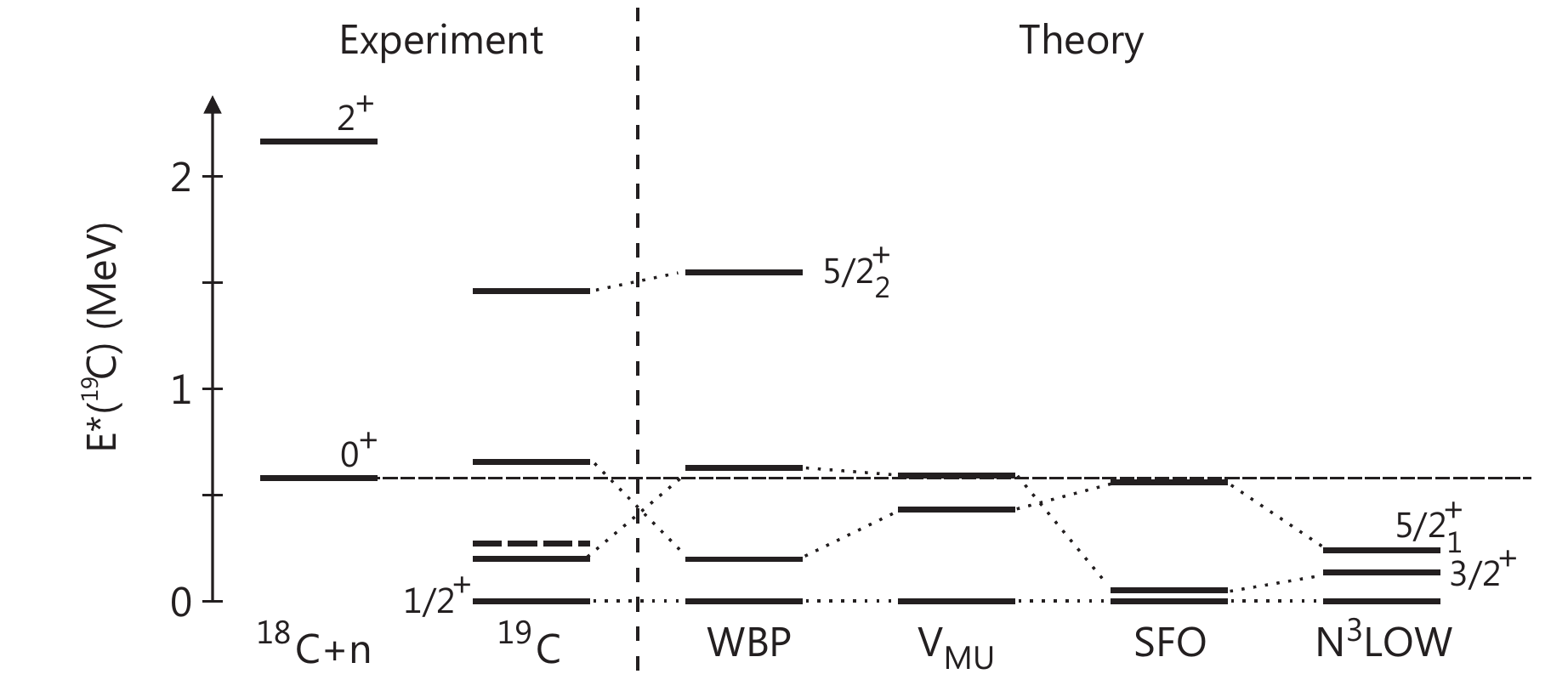}
        \end{center}
        \caption{Partial level schemes of \nuc{18}{C} and \nuc{19}{C}. The experimental levels of \nuc{19}{C} are compared to several shell model calculations. See text for details.}
        \label{fig:level}
    \end{figure}

We propose reassigning the first $5/2^+_1$ state from the 269(8)~keV bound state to the 653(95)~keV unbound state observed in the present experiment. The proposed experimental level scheme of \nuc{19}{C} relative to the ground and first excited state of \nuc{18}{C} is shown in Figure \ref{fig:level} and compared to several shell model calculations. The experimental data for the first excited $2^+$ state in \nuc{18}{C} come from Refs. \cite{Stanoiu2008,Fifield1982,Voss2012} and the first and second bound states in \nuc{19}{C} from Refs. \cite{Stanoiu2004}, \cite{Stanoiu2008} and \cite{Elekes2005}, respectively. The previously observed state at 269(8)~keV is shown by the dashed line. The two unbound $5/2^+$ states come from the present work and Ref. \cite{Satou2008}.

It should be mentioned that the present resonance might correspond to a high-lying state which decays via the first excited 2$^+$ state in \nuc{18}{C}. Although it is unlikely, it cannot be ruled out by the present data and $\gamma$-ray conicidence measurements would be necessary to exclude this possibility.

Finally, shell model calculations with the code NuShellX@MSU \cite{NuShellX} were performed in a truncated {\it s-p-sd-pf} model space with the standard WBP interaction. As shown in Figure \ref{fig:level}, this interaction reproduces the level spacing. However, the order of the $3/2^+$ and the $5/2^+_1$ states is reversed. Similarly, a recent calculation by Yuan \textit{et al.} using a newly constructed shell-model Hamiltonian developed from a monopole-based universal interaction ($V_{MU}$)  \cite{Yuan2012} does not reproduce the experimental proposed ordering of these states. However, the level ordering is reproduced by a different calculation presented by Yuan  \textit{et al.} in the same paper \cite{Yuan2012} using the SFO effective interaction \cite{Suzuki2003}. Calculations by Coraggio \textit{et al.} \cite{Coraggio2010}, who derived the single-particle energies and the residual two-body interaction of the effective shell-model Hamiltonian from the realistic chiral NN potential N$^3$LOW, also agree with the experimental level ordering but do not reproduce the measured level spacing.

\section{Conclusions}

A low-lying unbound state in \nuc{19}{C} was observed in one-proton (plus two neutron) removal reactions from a \nuc{22}{N} beam. The measured decay energy of 76(14)~keV corresponds to an excitation energy of  653(95)~keV. This state could correspond to the missing unbound $5/2^+_1$ state which had been proposed by one- and two-neutron removal reactions from \nuc{20}{C}. This assignment opens up the question of the spin and parity of the second bound excited state at 269(8)~keV which had previously been tentatively assigned to $5/2^+_1$.

\section*{Acknowledgments}

We would like to thank Robert Haring-Kaye and Sharon Stephenson for their valuable comments. This work was supported by the National Science Foundation under grants PHY-05-55488, PHY-05-55439, PHY-06-51627, PHY-06-06007, PHY-08-55456, PHY-09-69173, and PHY-11-02511.

    \bibliographystyle{nphys}

    %    \bibliography{/home/smosby/bibliography/nscl_c21c19}

\begin{thebibliography}{00}

%\bibitem{Baumann2012} T.Baumann, A.Spyrou, M.Thoennessen, Rep.Prog.Phys. 75, 036301 (2012)

\bibitem{Elekes2007} Z. Elekes {\it et al.}, Phys. Rev. Lett. {\bf 98} (2007) 102502
\bibitem{Schiller2007} A. Schiller {\it et al.}, Phys. Rev. Lett. {\bf 99} (2007) 112501
\bibitem{Hoffman2009} C.R. Hoffman {\it et al.}, Phys. Lett. B {\bf 672} (2009) 17
\bibitem{Hoffman2011} C.R. Hoffman {\it et al.}, Phys. Rev. C {\bf 83} (2011) 031303
\bibitem{Satou2012} Y. Satou {\it et al.}, Few Body Systems, DOI:10.1007/s00601-012-0377-3 (2012)
\bibitem{Tsoo2012} K. Tshoo {\it et al.}, Phys. Rev. Lett. {\bf 102} (2012) 022501
\bibitem{Lapoux2012} V. Lapoux {\it et al.}, Prog. Theor. Phys.(Kyoto), Suppl. {\bf 196} (2012) 111
\bibitem{Hoffman2008} C.R. Hoffman {\it et al.}, Phys. Rev. Lett. {\bf 100} (2008) 152502
\bibitem{Lunderberg2012} E. Lunderberg {\it et al.}, Phys. Rev. Lett. {\bf 108} (2012) 142503
\bibitem{Satou2008} Y. Satou {\it et al.}, Phys. Lett. B {\bf 660} (2008) 320
\bibitem{Bowman1974} J.D. Bowman, A.M. Poskanzer, R.G. Korteling, and G.W. Butler, Phys. Rev. C {\bf 9} (1974) 836
\bibitem{Thoennessen2012} M. Thoennessen, At. Data. Nucl. Data Tables {\bf 98} (2012) 43
\bibitem{Nakamura1999}  T. Nakamura {\it et al.}, Phys. Rev. Lett. {\bf 83} (1999) 1112
\bibitem{Maddalena2001} V. Maddalena {\it et al.}, Phys. Rev. C {\bf 63} (2001) 024613
\bibitem{Yuan2012} C. Yuan, T. Suzuki, T. Otsuka, F. Xu, and N. Tsunoda, Phys. Rev. C {\bf 85} (2012) 064324
\bibitem{Stanoiu2004} M. Stanoiu {\it et al.}, Eur. Phys. J. A {\bf 20} (2004) 95
\bibitem{Stanoiu2008} M. Stanoiu {\it et al.}, Phys. Rev. C {\bf 78} (2008) 034315
\bibitem{Elekes2005} Z. Elekes {\it et al.}, Phys. Lett. B {\bf 614} (2005) 174
\bibitem{Ozawa2011} A. Ozawa {\it et al.}, Phys. Rev. C {\bf 84}  (2011) 064315
\bibitem{Kobayashi2012} N. Kobayashi {\it et al.}, Phys. Rev. C {\bf 86} (2012) 054604
\bibitem{Morrissey2003} D.J. Morrissey {\it et al.}, Nuc. Instr. Meth. in Phys. B {\bf 204} (2003) 90
\bibitem{Mosby2012} S. Mosby {\it et al.}, Nucl. Phys. A, in press (2013)
\bibitem{Bird2005} M. Bird {\it et al.}, IEEE Trans. Appl. Supercond. {\bf 15} (2005) 1252
\bibitem{Baumann2005} T. Baumann {\it et al.}, Nuc. Instr. Meth. in Phys. A {\bf 543} (2005) 517
\bibitem{Christian2012} G. Christian {\it et al.}, Phys. Rev. C {\bf 85} (2012) 034327
\bibitem{Tostevin1999} J.A. Tostevin, J. Phys. G {\bf 25} (1999) 735
\bibitem{Bazin2003} D.Bazin {\it et al.} Phys. Rev. Lett. {\bf 91} (2003) 012501
\bibitem{Wang2012} M. Wang, G. Audi, A.H. Wapstra, F.G. Kondev, M. MacCormick, X. Xu, and B. Pfeiffer, Chin. Phys. C {\bf 36} (2012) 1603
\bibitem{Audi2003} G. Audi, A. Wapstra, and C. Thibault, Nucl. Phys. A, {\bf 729} (2003) 337
\bibitem{Mass2003} http://amdc.in2p3.fr/masstables/Ame2003/rct2.mas03
\bibitem{Fifield1982} L.K. Fifield {\it et al.}, Nucl. Phys. A {\bf 385} (1982) 505
\bibitem{Voss2012} P. Voss {\it et al.}, Phys. Rev. C {\bf 86} (2012) 011303
\bibitem{NuShellX} http://www.nscl.msu.edu/$\sim$brown/resources/resources.html
\bibitem{Suzuki2003} T. Suzuki, R. Fujimoto, and T. Otsuka, Phys. Rev. C {\bf 67} (2003) 044302
\bibitem{Coraggio2010} L. Coraggio, A. Covello, A. Gargano, and N. Itaco, Phys. Rev. C {\bf 81} (2010) 064303



\end{thebibliography}
\end{document}